\documentclass[fleqn,twoside]{article}
\usepackage{espcrc2}

\readRCS
$Id: espcrc2.tex,v 1.2 2004/02/24 11:22:11 spepping Exp $
\usepackage{graphicx}
\usepackage[figuresright]{rotating}

\title{CVD Diamonds in the BaBar Radiation Monitoring System}
\author{M. Bruinsma\address[UCI]{University of California at Irvine, Irvine, CA 92697}\thanks{email: bruinsma@slac.stanford.edu},%
        P. Burchat\address[SU]{Stanford University, Stanford, CA 94305-4060}, 
	A.J. Edwards\addressmark[SU], %
       	H. Kagan\address[OSU]{Ohio State University, Columbus, OH 43210}, 
	R. Kass\addressmark[OSU], 
	D. Kirkby\addressmark[UCI] 
	and 
	B.A. Petersen\addressmark[SU] 
             } 

\begin{document}

\begin{abstract}
To prevent excessive radiation damage to its Silicon Vertex Tracker, the BaBar experiment at SLAC
uses a radiation monitoring and protection system that triggers a beam abort whenever radiation
levels are anomalously  high. The existing system, which employs large area Si PIN diodes as radiation sensors,
 has become increasingly difficult to operate due to radiation damage.
 
We have studied CVD diamond sensors as a potential alternative for these silicon sensors.
Two diamond sensors have been routinely used since their installation in the Vertex Tracker in August 2002.
The experience with these sensors and a variety of tests in the laboratory have
shown
CVD diamonds to be a viable solution for dosimetry in high radiation environments.
However, our studies have also revealed surprising side-effects.

\vspace{1pc}
\end{abstract}

\maketitle

\section{Introduction}
The BaBar experiment\cite{BaBar} at the Stanford Linear Accelerator Center (SLAC)
employs a radiation monitoring and protection system
SVTRAD\cite{SVTRAD} to safeguard  the Silicon Vertex Tracker (SVT)\cite{SVT} from
excessive radiation damage. The SVTRAD system uses 12 Hamamatsu S3590-08 silicon PIN diodes with dimensions
1$\times$1~cm$^{2}$ $\times$ 300~$\mu$m, arranged in two rings and mounted close to the readout electronics 
of the Layer-1 SVT modules. The diodes are reverse biased at 50V and are connected to a custom-built, DC-coupled read-out board.
This board measures the total current flowing through the sensor, which consists of a dominant leakage current and
a small radiation-induced current. The readout board  regularly calibrates the 
baseline leakage current when beams are absent and corrects for temperature fluctuations measured by 
thermistors installed close to the PIN diodes. 
The read-out board continuously subtracts the estimated leakage-current from the total current to obtain the
radiation-induced component, and issues a beam abort when the radiation currents are above predefined threshold values.

SVTRAD has been in active operation since the start of BaBar in 1999 and has issued on average 2-3 beam aborts
per day. While successful so far in keeping radiation damage to the SVT within acceptable limits, the system has 
become increasingly difficult to operate because of radiation damage to the silicon 
sensors. Some of the sensors have accumulated a dose of approximately 2 Mrad, resulting in leakage currents exceeding
 3$~\mu A$. The precision on the estimated pedestal is approximately 0.1\% of the leakage current, resulting in 
a systematic uncertainty on radiation currents of a few nA - comparable in magnitude to average radiation levels.
Without intervention, the system would become inoperable within one or two years due to further radiation damage. 
We therefore searched for sensor materials that are more radiation hard than silicon. The most
promising material was CVD diamond.

\section{CVD diamonds}
The use of polycrystalline CVD diamond as a radiation - or particle detector 
has been extensively studied by the RD42 collaboration\cite{RD42a,RD42b}. Over the last 10 years, the
sensitivity to radiation has improved by almost an order of magnitude, with charge-collection-distances now
reaching levels of 250~$\mu$m. 
Polycrystalline CVD (pCVD) diamond sensors are grown on silicon substrates using hydrogen and methane 
as source gases. An artifact of the production process is that diamonds exhibit a columnar growth structure, with
grain dimensions of typically 10-100~$\mu$m. Much of the electrical properties of pCVD diamonds are determined by the grain boundaries, where
graphite and other impurities are accumulated. These impurities lower the resistivity and 
act as possible traps of drifting charges, leading to a loss of sensitivity to radiation.

When operated as a radiation sensor, a high voltage is applied across the electrodes on either side of the CVD diamond wafer.
Electron-hole pairs created by incident ionizing particles drift apart in the electrical field, leading to a current flow that
is proportional to the dose rate. Maximum charge collection for pCVD sensors is achieved at an 
electric field of typically 1V/$\mu$m. 

pCVD diamond sensors do not reach their full efficiency immediately, but only after they are fully 'pumped' by radiation. In this process,
the charge traps in the grain boundaries are filled so that they no longer act as a source of inefficiency.

An important benefit of diamond over silicon as a radiation sensor material is its radiation hardness. 
Crucially important for the application as a DC coupled dosimeter in BaBar is that leakage currents are virtually absent 
(few pA) and do not increase with accumulated dose.

\section{CVD diamonds in BaBar}
To determine whether diamond sensors present a viable alternative to PIN diodes,
two CVD diamond sensors of 1x1~cm$^2$ area and 0.5~mm thickness have been installed inside the BaBar SVT in August 2002.
They were placed in the horizontal plane of the detector, on either side of the  beam pipe, approximately 15~cm from the $e^{+}e^{-}$ interaction point
and 5~cm from the beam line. Much of the radiation in BaBar is concentrated in this horizontal plane, since
off-track beam particles are swept in the horizontal direction by the final focusing magnet. The dose rate during stable beam operation is 
typically 25~mrad/s in this horizontal plane, and approximately 5~mrad/s elsewhere at the same radius.

The collection distance of both sensors has been measured to be approximately 200~$\mu$m at 500~V. Both diamonds are
biased at 500~V to ensure maximum sensitivity. The sensitivity of this diamond sensor has been measured to be approximately 
100pC/mrad, compared to 200pC/mrad for the Si PIN diodes.  Ohmic contacts of 0.9x0.9~cm$^2$ are applied to
either side and are soldered to coax cables with Indium solder, electrically shielded and insulated.

The currents in the sensors are measured with the same type of read-out board that is used for the Si PIN diodes. The readout is
 integrated in the BaBar online monitoring system.

\section{Operational Experience}
\begin{figure}[b!]
\includegraphics[width=0.5\textwidth]{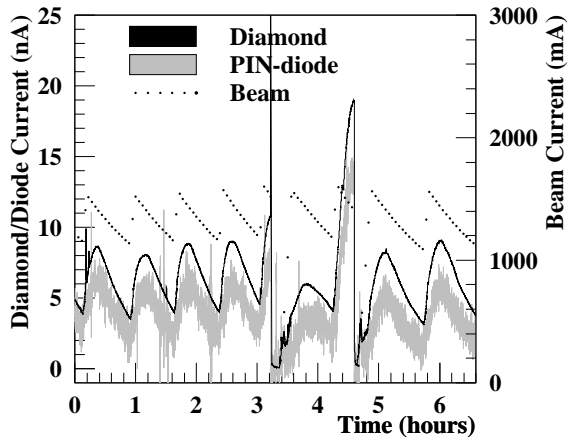}
\vspace{-1.5cm}
\caption{Radiation-induced currents in diamond sensor and Si PIN diode during typical operation of the accelerator. The upper curve shows 
the current of the electron beam, which gradually decreases between injections. The burst of radiation at $t=3.2$~hours, which caused the beams 
to be dumped, is seen in both sensors.}
\label{fig:Tuesday_ler_BW}
\end{figure}

The sensors have been in routine operation for almost two years, presenting the longest continuous operation of diamond sensors
and the first such use in a running High-Energy Physics experiment. The current in one of the two diamond sensors during typical beam conditions
is shown in figure \ref{fig:Tuesday_ler_BW}, along with the radiation levels as measured by one of the 12 PIN diodes. It can be clearly seen that
variations in dose rates due to differing beam currents are well tracked by both sensors. The diamond sensor, however, gives a more precise 
measurement since it is not affected by temperature fluctuations. The correlation between the radiation levels and the current in the diamond 
sensor is excellent.
\begin{figure*}[t]
\includegraphics[width=0.49\textwidth]{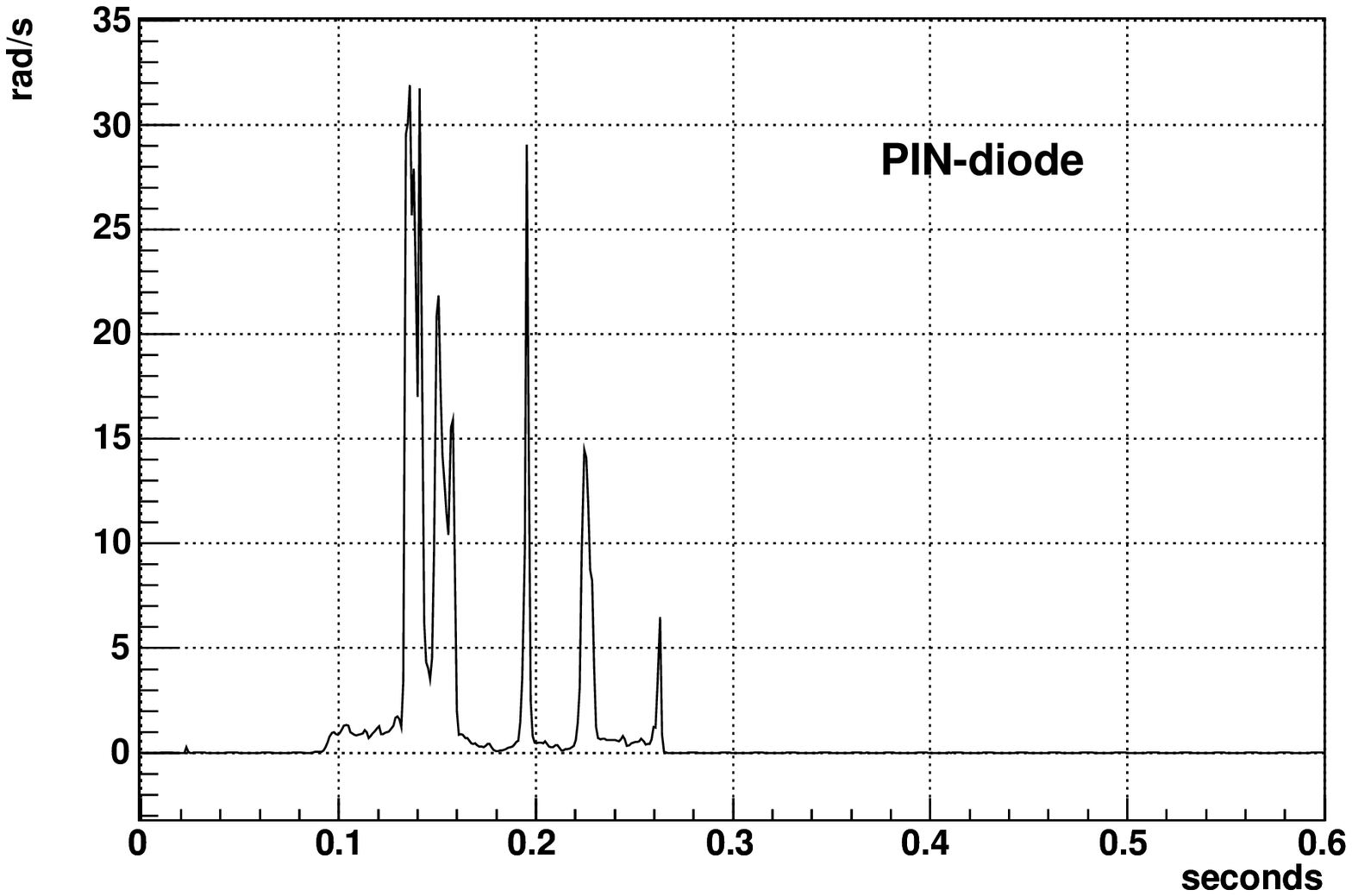}
\includegraphics[width=0.49\textwidth]{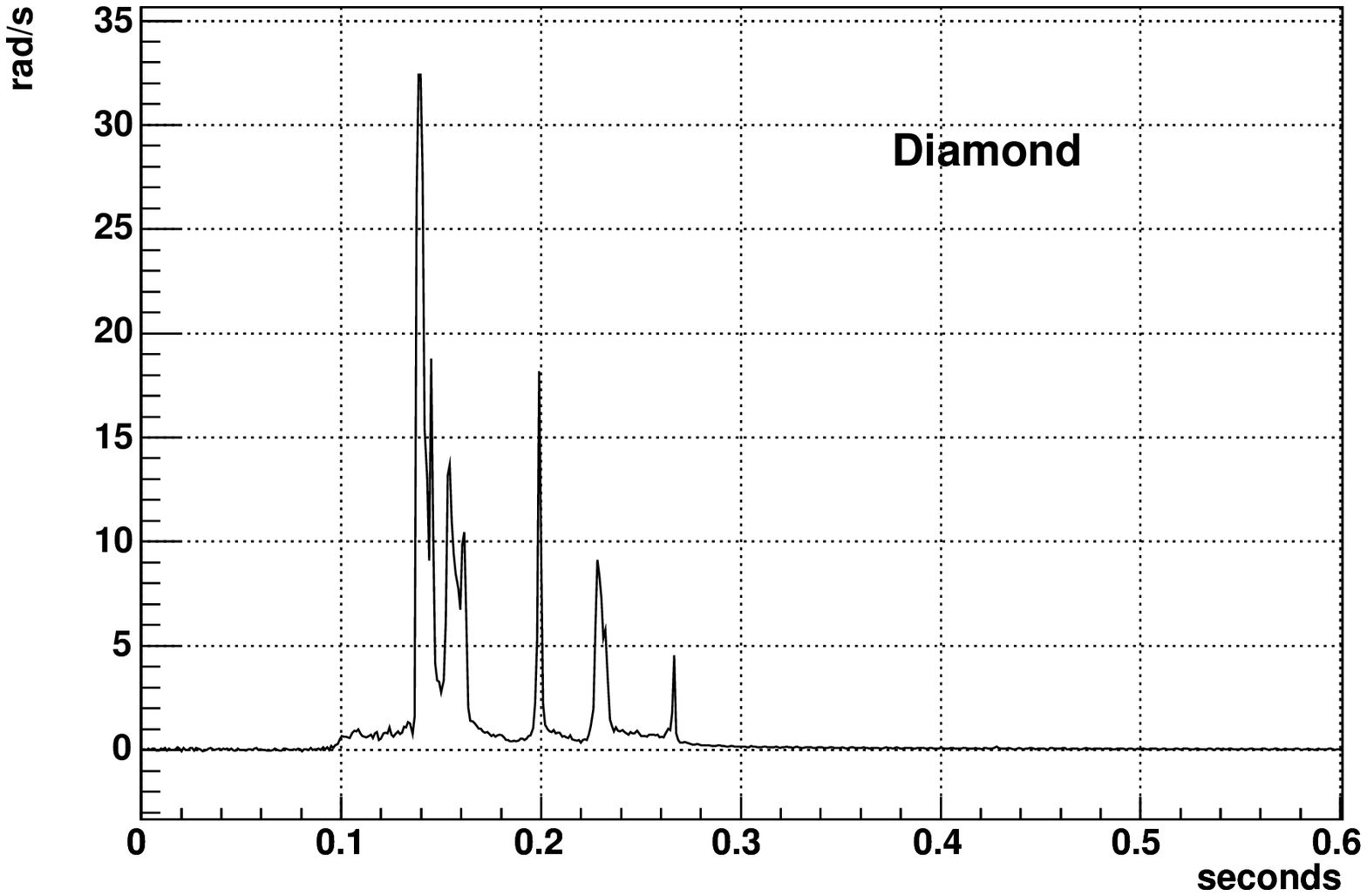}
\vspace{-1cm}
\caption{Response of PIN diode and diamond sensor during bursts of radiation causing a beam abort.}
\label{fig:aborts}
\end{figure*}
The SVTRAD system stores a 10 second long snap-shot of the radiation levels prior to each beam abort.
Figure \ref{fig:aborts} shows that the response of the diamond sensor on such small time scales is also 
well correlated to the response in other sensor. In a dedicated tests with a fast custom-built amplifier 
we have measured the response time on spikes of radiation to be less than 20~ns, in agreement with and 
limited by the bandwidth of the amplifier. This is comfortably within the limit of 100~$\mu$s required for the SVTRAD system.

Up to now, the diamond sensors have received over 500~krad of integrated radiation dose. No loss of sensitivity and no 
increase of leakage current has been observed so far.

\section{Side-effects}
\begin{figure}[bt]
\includegraphics[width=0.5\textwidth]{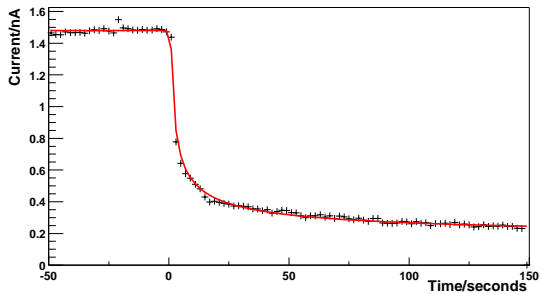}
\vspace{-1.5cm}
\caption{remnant current in the diamond after loss of radiation source. The signal exhibits a non-exponential
tail that is fit to a $1/\sqrt{t}$ functional dependence (curve).}
\label{fig:tails}
\end{figure}

We have observed notable side-effects related to the stability of the
current in pCVD sensors.  One is the presence of non-exponential tails after the sudden drop in 
radiation levels, as illustrated in figure \ref{fig:tails}. This 'after-glow' is visible on time-scales from 
milliseconds to minutes, and has been reported elsewhere\cite{tails1,tails2,tails3}.
It could be explained by the thermal evaporation of charge stored in shallow traps inside the diamond bulk.
The non-exponential nature of the tails points to the presence of multiple traps with different energies.
A continuum of traps of different relaxation times $\tau$ and occupation levels 
$N(\tau) \propto 1/\tau^{3/2}$ leads to a $1/\sqrt{t}$ dependence, which gives a good description of the data.
 

Another side-effect is the occurrence of erratic dark currents, which were discovered when the
1.5~T magnetic field that immerses the BaBar detector temporarily switched off. Right after the magnetic field 
disappeared, dark currents started rising from a few pA to 100~nA for one diamond sensor and to 4~nA for the other,
both fluctuating by up to 50\%. When the magnetic field reappeared the dark currents vanished.

We have carried out an extensive test program to reproduce this effect in the lab and to elucidate the nature of the
erratic dark currents. For these test we used four new sensors of the same dimension as the sensors installed. 
Two of four sensors had the electrodes segmented in 2x2 pads and all sensors were equipped with a guard ring, giving a total
of 10+4 channels on four sensors. Furthermore,
half of the sensors were metalized with Al to study the effect of the Schottky barrier.

Before irradiating the sensors they were biased at 500~V for three weeks to see whether the dark currents could appear spontaneously.
We found that erratic dark currents appeared in 7 of the 14 channels after a period ranging between 2 days
and two weeks. Since no anomalously high currents were present at the guard ring, and since a single pad could show erratic dark currents
while the other three pads were still quiet, we conclude it to be a local effect.

The sensors were subsequently irradiated in a $^{\rm 60}$Co source at different bias voltages and at different dose rates. 
We found no erratic dark currents in any of the sensors when biased at 100~V for a week. In contrast, erratic dark currents 
appeared in each of the 14 channels when biased at 500~V. The currents were orders of magnitude higher than before irradiation
and persisted outside the $^{\rm 60}$Co source. No differences have been observed between the two different metalizations.

\begin{figure}[bt]
\includegraphics[width=0.45\textwidth]{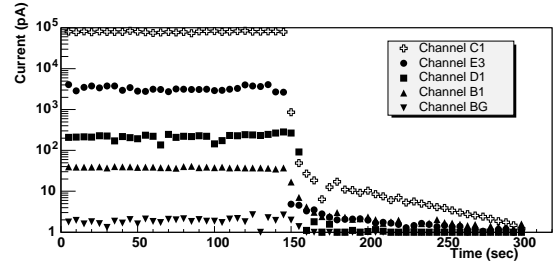}
\vspace{-1.0cm}
\caption{Suppression of erratic dark currents in a magnetic field. At $t=150$~s a 1.5T magnet was switched on,
resulting in the elimination of erratic dark currents in all of the sensors.}
\label{fig:magsuprr}
\end{figure}

We have verified that a sufficiently high ($>1$T) magnetic field, perpendicular to the electric field inside the sensor, always
eliminates the erratic dark currents, as shown in figure \ref{fig:magsuprr}.

Indications are that the (increase of) erratic dark currents due to irradiation can be undone by heating the sensor to 400~$^\circ$C,
a temperature at which charge stored in deep traps is thermally evaporated. 
At temperatures below 100~$^\circ$C, the erratic dark currents increase by approximately 4\%/$^\circ$C.

\section{Conclusions and outlook}
In almost two years of operation, pCVD diamond radiation sensors have proven to be a viable alternative 
to silicon PIN diodes for radiation monitoring in BaBar. Although intended as a feasibility study, the two diamond sensors 
installed already play a vital role in BaBar, giving a 
more precise and stable measure of the dose rate than the PIN diodes. Still, there are important 
side-effects in diamonds that deserve particular attention. In the summer of 2005 we plan to install 12 new diamond sensors
to replace all of the PIN diodes.

We thank the RD42 collaboration and Element Six for providing the CVD diamond sensors.

\end{document}